# Channel Capacity Limitations versus Hardware Implementation for UWB Impulse Radio Communications

Aubin Lecointre, Daniela Dragomirescu, and Robert Plana.
University of Toulouse
LAAS-CNRS
7, Av du Colonel Roche
31077 Toulouse cedex 4, France
{alecoint,daniela,plana}@laas.fr

*Abstract*—Starting from the Shannon channel capacity, we propose an IR-UWB channel capacity based on the delay spread for multipath time variant channels. This IR-UWB channel capacity is obtained from the no ISI (Inter Symbol Interference) assumption and for binary modulations. The impact of the kind of implementation is considered on the IR-UWB channel capacity. This study is lead for mixed and mostly digital implementation. The key parameters and theirs impacts on the channel capacity are exposed in each case: the data converters for mostly digital implementations and the pulse generator capabilities for mixed implementations. Finally, these two implementations are compared from a data rate point of view. Their behaviors regarding an increase of the operating frequency are also studied.

## I. Introduction

THIS paper proposes a study of the IR-UWB (Impulse Radio Ultra WideBand) channel capacity, by using a new expression obtained from Shannon capacity [1], which takes in consideration the kind of implementation. IR-UWB could be designed in a mostly digital radio way [2] or in a classical mixed way. This study exposes the architecture key points and their importance from a high data rate point of view. By proposing to merge the IR-UWB channel capacity study and implementation considerations, we are able to specify the dimensioning element for each kind of architecture. Achievable data rate values, for mixed and mostly digital implementation, are also obtained thanks to the IR-UWB channel capacity. In this article only the binary modulations will be considered, in order to emphasize the IR-UWB simplicity behaviour.

This paper is laid out as follow, at first, in Section II, we present a general and very simple form of the IR-UWB channel capacity based on delay spread, for binary modulations over time variant multipath channels and under the no ISI (Inter Symbol Interference) assumption. In the Section III, we exhibit the IR-UWB channel capacity for a mostly digital implementation, while Section IV is dedicated to the classical mixed implementation. At last Section V is devoted to the conclusion.

## II. IR-UWB Channel Capacity

From the Shannon capacity, we expose the IR-UWB channel capacity (1) by considering for the channel temporal resolution with the delay spread and for the channel amplitude resolution thanks to the Shannon formula. The IR-UWB channel capacity proposed here is defined for an absence of ISI. By dealing with the channel delay spread, which is among the most of important parameters for modelling IR-UWB channel, we can defined the IR-UWB channel capacity for time-variant multipath channels. Fig. 1 illustrates the IR-UWB configuration regarding symbol, pulse duration ($T_p$) and delay spread ($d_{RMS}$) for achieving the largest data rate, and thus determine the channel capacity.

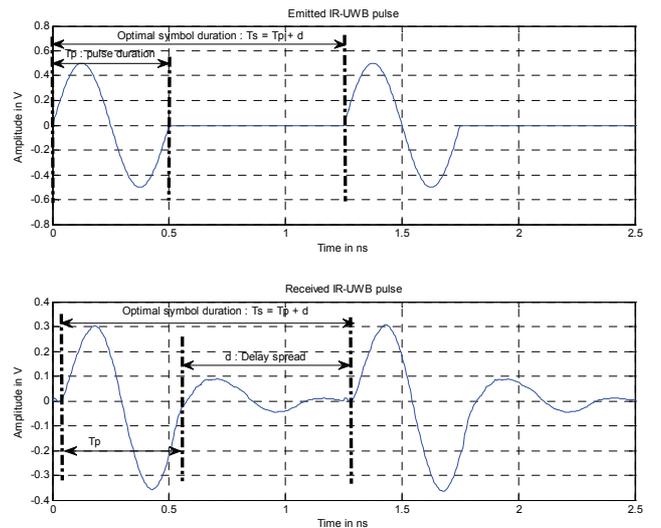

Fig. 1 − Illustration of the configuration of the symbol duration, in function of the pulse duration and the channel delay spread, for achieving the better data rate.

This work was supported by the the French research national agency ANR, under the RadioSOC project (n°JC05-60832). A. Lecointre thanks to DGA, "Délégation Générale pour l'Armement", for PhD funding.

$$C_{IR-UWB}(bits/s) = \frac{1}{T_p + d_{RMS}} \times \frac{1}{2} \times \log_2(1 + SNR) \quad (1)$$

With $T_p$ the IR-UWB pulse duration, $B$ the bandwidth of the IR-UWB signal, $d_{RMS}$ the root mean square channel delay spread and $SNR$ the signal-to-noise ratio.

Equation (2) is a particular case of (1). Equation (2) defines the IR-UWB channel capacity only for the binary modulations. A 3 dB signal-to-noise ratio (SNR) is required for binary modulations at the receiver.

$$C_{IR-UWB binary\_mod}(bits/s) = \frac{1}{T_p + d_{RMS}} = \frac{1}{\frac{1}{B} + d_{RMS}} \quad (2)$$

With $T_p$ the IR-UWB pulse duration, $B$ the bandwidth of the IR-UWB signal, and $d_{RMS}$ the root mean square channel delay spread.

The IR-UWB channel capacity for binary modulations, versus the channel bandwidth, is represented on fig. 2, for three distinct channel delay spreads.

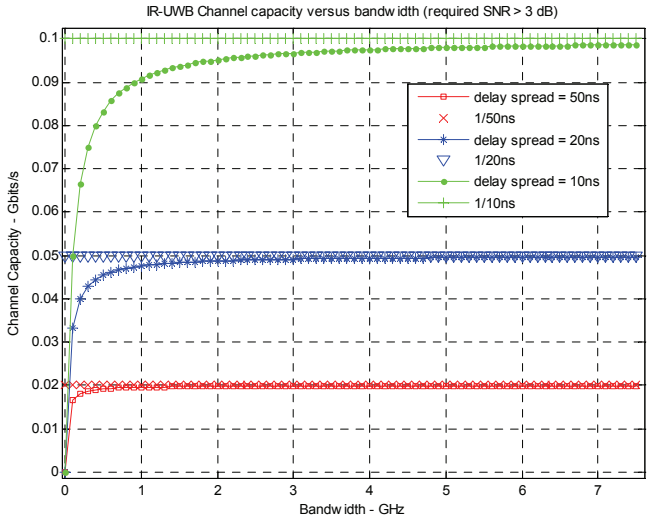

Fig. 2 – Illustration of the IR-UWB channel capacity, for binary modulations, on the UWB bandwidth, for three channel delay spreads.

The IR-UWB channel capacity analysis exposes that the most predominant parameter is the channel delay spread. Indeed, this latter will determine the limit of the channel capacity whatever the channel bandwidth increase. Thus the channel capacity is limited by an asymptote at $1/d_{RMS}$. As a result a decrease of the IR-UWB pulse duration, which requires higher circuit operating frequency, will not permit to increase efficiently the IR-UWB channel capacity.

### III. MOSTLY DIGITAL IMPLEMENTATION LIMITATIONS ON IR-UWB CHANNEL CAPACITY

If we consider a mostly digital implementation for IR-UWB as described in fig. 3, the IR-UWB channel capacity is now dimensioned by the channel delay spread and also by the key points of this architecture. They are analog-to-digital converters (ADC), digital-to-analog converters (DAC) and the digital part dedicated for the digital signal processing: FPGA (Field Programmable Gate Array) or ASIC (Application Specific Integrated Circuit).

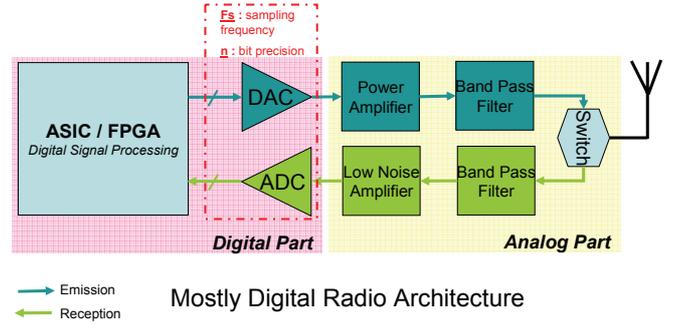

Fig. 3 – The mostly digital radio implementation for an IR-UWB radio. Illustration of ADC/DAC as the key component.

The performances of this part will dimension the performances of the transceiver and also the achievable data rate. That's why we propose to determine the IR-UWB channel capacity (3), in the binary modulation case, for a mostly digital radio in function of the sampling frequency ($F_s$) used in ADC/DAC.

$$C_{IR-UWB\ MOSTLY-DIGITAL}(bits/s) = \frac{1}{\frac{n_{sampling}}{F_s} + d_{RMS}} \quad (3)$$

With $n_{sampling}$ the sampling factor, i.e. the ratio between the sampling frequency and the analog signal maximum frequency (the inverse of the IR-UWB pulse duration); $F_s$ the sampling frequency of the data converters, and $d_{RMS}$ the RMS channel delay spread.

The digital circuit frequency is not a dimensional element, since thanks to techniques such as time interleaved ADC, the digital signal processing is done at a lower frequency. It requires a parallelization of the processing and a retiming algorithm at the output of data converters [3] [4]. As a result only sampling frequency of the data converter has to be taken into consideration in the evaluation of the channel capacity, since it's the most dimensioning parameter.

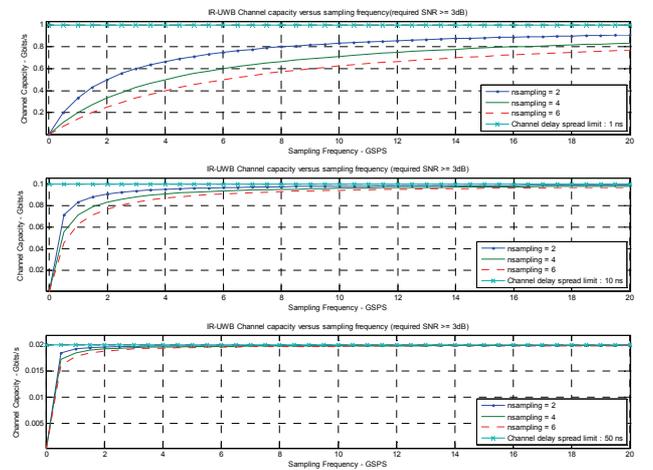

Fig. 4 – IR-UWB channel capacity for mostly digital radio implementation for three distinct channel delay spreads. Impact of sampling frequency and sampling factor is exposed.

Figure 4 exposes the IR-UWB channel capacity for binary modulations versus sampling frequency for different channel delay spreads and $n_{sampling}$ factors.

The binary modulation framework is reinforced, in mostly digital implementation, by the flash ADC capability. Indeed, flash converters, for power consumption and surface reasons can achieve a high sampling frequency or a high bit precision but not both [5]. Besides the binary modulations require a fewer bit precision than the M-ary modulations. Due to the IR-UWB bandwidth we are forced to consider high speed data converters and thus binary modulations for respecting the flash converters capabilities.

From fig. 4, an analysis of the IR-UWB channel capacity regarding the delay spread, the sampling frequency and the sampling factor can be done. For mostly digital implementation, the delay spread remains the most important limitation. It defines an asymptote at $1/d_{RMS}$. The higher the sampling frequency is, the higher the achievable data rate is. Low delay spread channels require higher sampling frequencies, for yielding the channel capacity, than high delay spread channels. For illustrating this, we propose, on fig. 5, to expose the percentage of the maximum channel capacity versus the sampling frequency. With this representation, this previous statement can be easily illustrated. If we consider a 90% threshold, for low channel delay spreads it will be reached for a higher sampling frequency than the sampling frequency required, in high channel delay spread case for achieving this 90% threshold.

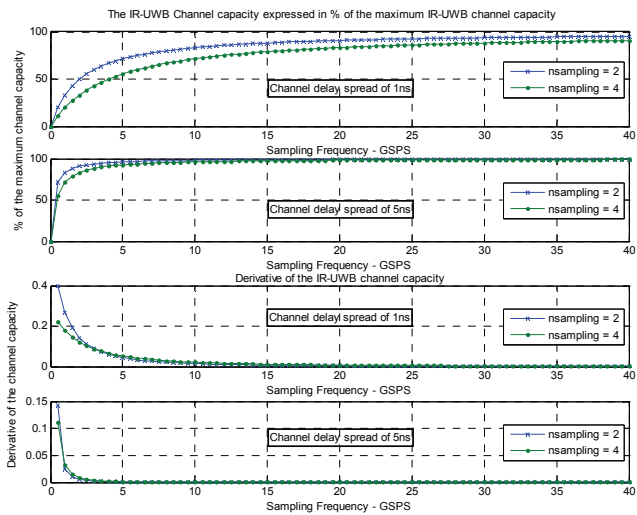

Fig. 5 – IR-UWB channel capacity for mostly digital radio implementation expresses in percent of the maximum capacity. The derivative of the channel capacity is also exposed.

Fig. 5 allows to visualize that the achievable data rate difference for two successive low sampling frequencies is larger than for two successive higher sampling frequencies. This is due to the presence of the delay spread asymptote. For high sampling frequencies the channel capacity is nearer to the asymptote than for low sampling frequencies. This behaviour could be illustrated by visualizing the derivative of the IR-UWB channel capacity on fig. 5. This latter exposes that for very high sampling frequencies the derivative tends toward zero and thus an increase of the sampling frequency is useless and especially inefficient, in a high data rate context. Nevertheless, state-of-the art sampling frequencies are relatively small. Thus, as the channel capacity derivative shows, an increase of the sampling frequency for low sampling frequencies is efficient regarding the data rate.

Concerning the sampling factor, the higher it is, the lower the channel capacity is. An increase of the sampling factor implies a decrease of the channel capacity, but it permits to achieve better performances regarding the bit error rate (BER) and, for example, the synchronization precision [6]. Thus, there is a data rate versus performance trade-off. Fig. 4 exposes also that the gain of low sampling factors, for achieving high data rates, decrease with an increase of the sampling frequency. This is due to the delay spread asymptote behaviour of the IR-UWB channel capacity.

Achievable data rate values for the IR-UWB mostly digital implementations can be determined by using the state of the art regarding the data converter performances [7] and the sampling factor. This later has generally a value of four [6], this value is adapted for achieving a good balance regarding correlation and synchronization performances while minimizing the power consumption. For obtaining realistic IR-UWB channel delay spreads, the IEEE 802.15.4a channel model is used. Table I and II summarize these information.

TABLE I
STATE OF THE ART A/D CONVERTERS [7]

| Designer | Year | Sampling Frequency | Bit Precision | Dissipated Power (W) | Ref. |
|---|---|---|---|---|---|
| *State of the art A/D converters* | | | | | |
| W. Yang et al. | 2001 | 75 MSPS | 14 | 0,35 | [8] |
| Y. Akazawa et al. | 1987 | 400 MSPS | 8 | | [9] |
| I. Mehr and L. Singer | 1999 | 500 MSPS | 6 | | [10] |
| HRL Labs | 1988 | 1 GSPS | 4 | 0.1 | [7] |
| IERU | 1988 | 1 GSPS | 4 | 2,4 | [7] |
| Fraunhofer & TriQuint | 1992 | 1 GSPS | 5 | 3,4 | [7] |
| Signal Processing Tech | 1995 | 1 GSPS | 8 | 5,5 | [7] |
| Raytheon | 1989 | 1.20 GSPS | 5 | 3 | [7] |
| TRW | 1996 | 1,75 GSPS | 8 | | [7] |
| Rockwell | 1995 | 2 GSPS | 8 | 5,3 | [7] |
| T. Wakimoto et al. | 1988 | 2 GSPS | 6 | | [11] |
| LEPA | 1986 | 3 GSPS | 4 | 0,15 | [7] |
| S. Park et al. | 2006 | 4 GSPS | 4 | 0,53 | [12] |
| HP & Rockwell | 1994 | 4 GSPS | 6 | 5,7 | [7] |
| HP | 1991 | 4 GPSS | 8 | 39 | [7] |
| HRL Labs | 1996 | 8 GSPS | 3 | 3,5 | [7] |
| J. Lee et al. | 2003 | 10 GSPS | 5 | | [13] |

TABLE II
A/D CONVERTERS AVAILABLE ON THE MARKET.

| Designer | Sampling Frequency | Bit Precision | Dissipated Power (W) |
|---|---|---|---|
| *Available on market A/D converters* | | | |
| Texas Instrument | 210 MSPS | 12 | 1,23 |
| Analog Device | 400 MSPS | 12 | 6,8 |
| Texas Instrument | 500 MSPS | 12 | 2,25 |
| e2v | 500 MSPS | 12 | 2,3 |
| e2v | 500 MSPS | 8 | 1,4 |
| National Semiconductor | 500 MSPS | 8 | 0,8 |
| Maxim | 600 MSPS | 8 | |
| Maxim | 1 GSPS | 8 | |
| National Semiconductor | 1 GSPS | 8 | 1,2 |
| National Semiconductor | 1,5 GSPS | 8 | 1,5 |
| Maxim | 1,5 GSPS | 8 | |
| e2v | 2 GSPS | 10 | 4,6 |
| e2v | 2,2 GSPS | 10 | 4,2 |
| Maxim | 2,2 GSPS | 8 | |
| National Semiconductor | 3 GSPS | 8 | 1,6 |
| e2v | 5 GSPS | 8 | 3,9 |

TABLE III
RMS CHANNEL DELAY SPREAD FOR IEEE 802.15.4A UWB CHANNEL BETWEEN 2 AND 10 GHZ.

| Environement | RMS delay spread (ns) |
|---|---|
| Residential LOS | 17 |
| Residential NLOS | 19 |
| Office LOS | 10 |
| Office NLOS | 13 |
| Outdoor LOS | 28 |
| Outdoor NLOS | 78 |
| Industrial LOS | 9 |
| Industrial NLOS | 89 |
| Open Outdoor NLOS | 21 |

Fig. 6 and table IV present these achievable data rate values if we consider state of the art components and realistic IR-UWB channels.

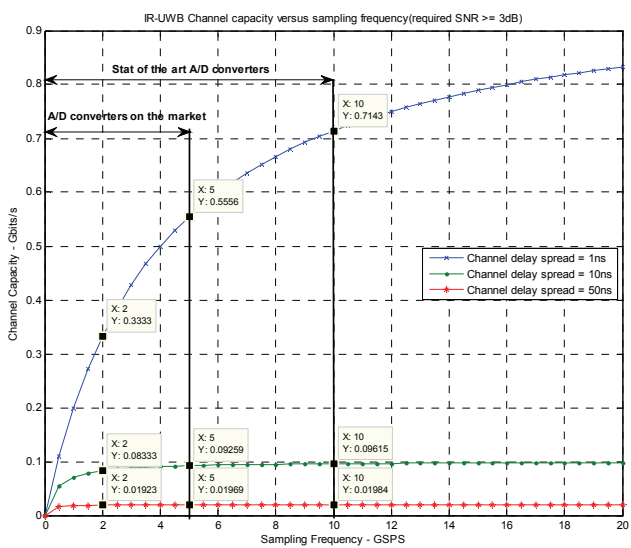

Fig. 6 – The IR-UWB channel capacity, in mostly digital cases, for state of the art and available on market data converters.

TABLE IV
ACHIEVABLE DATA RATE VALUES FOR MOSTLY DIGITAL IMPLEMENTATIONS, IN FUNCUTION OF REALISTIC CHANNEL DELAY SPREADS AND DATA CONVERTERS PERFORMANCES.

| Environement | RMS delay spread (ns) | Sampling Freq. of A/D (GSPS) | Sampling Factor | Channel Capacity (Mbits/s) |
|---|---|---|---|---|
| Residential LOS | 17 | 2,00 | 4,00 | 52,63157895 |
| Residential LOS | 17 | 5,00 | 4,00 | 56,17977528 |
| Residential LOS | 17 | 10,00 | 4,00 | 57,47126437 |
| Industrial LOS | 9 | 2,00 | 4,00 | 90,90909091 |
| Industrial LOS | 9 | 5,00 | 4,00 | 102,0408163 |
| Industrial LOS | 9 | 10,00 | 4,00 | 106,3829787 |
| Industrial NLOS | 89 | 2,00 | 4,00 | 10,98901099 |
| Industrial NLOS | 89 | 5,00 | 4,00 | 11,13585746 |
| Industrial NLOS | 89 | 10,00 | 4,00 | 11,18568233 |

For example for the industrial LOS channel a data rate of 90 Mbits/s can be attained by using a 2GSPS ADC converter (available on market) with a sampling factor of 4. The no ISI assumption and the binary modulations are again considered.

From the table IV and fig. 6, we can see that the data converter performance ($F_s$) is important, as a dimensioning factor, only for low channel delay spreads, regarding the IR-UWB channel capacity for binary modulations. As long as the channel delay spread is large, whatever the sampling frequency, the data rate can't be increased in a significantly manner (for binary modulations). However, the smaller the channel delay spread is, the larger the required sampling frequency has to be for yielding the channel capacity. In this case, when the channel delay spread is small, it is the data converters performances that limit the channel capacity.

## IV. IR-UWB CHANNEL CAPACITY FOR MIXED IMPLEMENTATION

With IR-UWB transceivers with the classical mixed implementations, the data converters performances are less preponderant than in the mostly digital radio case, since the converters are far from the antenna. Direct synthesis no longer exists in the mixed implementations. With this well-known technique of implementation the demodulation and the front-end is done in an analog way, while the digital signal processing is done in a digital way. Fig. 7 exposes one widely use architecture, among the mixed implementations.

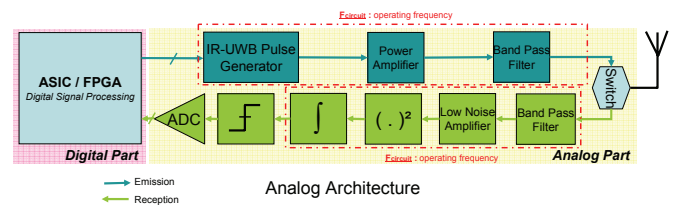

Fig. 7 – Illustration of an IR-UWB mixed architecture. The importance of the ADC is reduced in comparison with mostly digital implementation.

With the mixed implementations, the key parameter is the pulse generator and its ability to generate very short pulse duration at the emitter side. At the receiver side, the limitation is the operating frequency of circuits. Thus we consider in our expression of the IR-UWB channel capacity for binary modulations (4), only the most constraining frequency, i.e. the minimum one. Note that generally the analog operating frequencies are drastically greater than the sampling frequencies of data converters.

$$C_{IR-UWB\ ANALOG}(bits/s) = \frac{1}{\frac{1}{F_{circuit}} + d_{RMS}} \quad (4)$$

Where $F_{circuit}$ is the minimal operating frequency of the transceiver among all the analog circuits at the emitter and the receiver side; and $d_{RMS}$ is the RMS channel delay spread of the channel.

Fig. 8 illustrates the IR-UWB capacity for binary modulation and mixed implementation in function of the operating frequency of the circuit. Fig. 8 considers three channel delay spreads: 1; 5; and 10 ns.

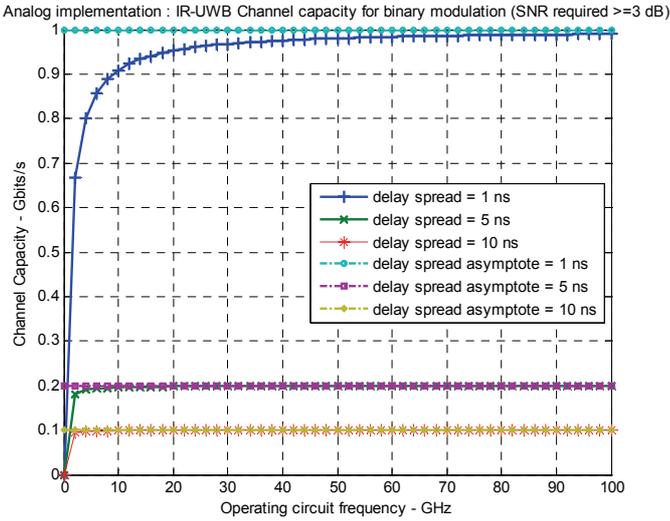

Fig. 8 – IR-UWB channel capacity in the case of binary modulations and mixed implementations.

Fig. 8 shows that the maximum IR-UWB channel capacity (the $1/d_{RMS}$ asymptote) is reached quickly when $F_{circuit}$ increase. This fact arrives more quickly in the case of high channel delay spreads than in low channel delay spreads. Typically, in a 10 ns channel delay and for an operating frequency of 5 GHz will give the same data rate capability than a 60GHz operating frequency. Thus, as in the mostly digital case, in the mixed implementation the channel delay spread is the preponderant dimensioning parameter. In the case of binary modulations and mono-band schemes, because of the relatively high channel delay spread in UWB realistic scenarios, the increase of the operating frequency is useless. Fig. 9 uses the derivative of the channel capacity for illustrating the ineffective of using high operating frequencies.

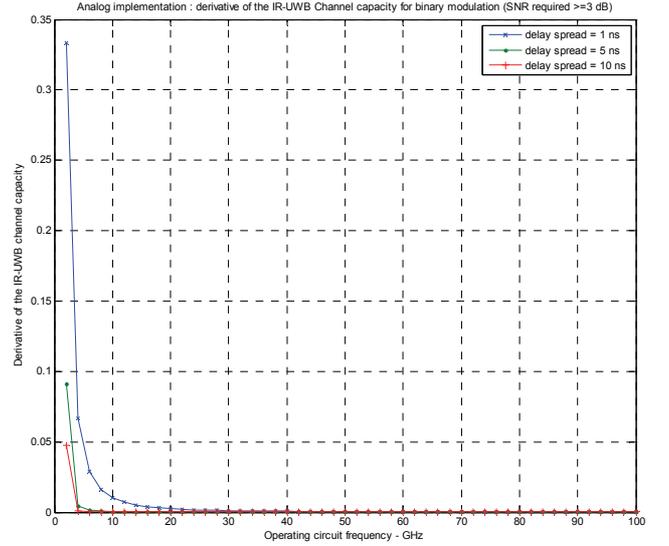

Fig. 9 – Derivative of the IR-UWB channel capacity for binary modulations and mixed implementations for three channel delay spreads.

Thanks to the analysis of the IR-UWB channel capacity, from a high data rate point of view, using an mixed implementation is interesting since it could permit to easier achieve high data rate, than in a mostly digital case, by using M-ary modulations. In mostly digital case, M-ary modulations are very difficult due to the antagonism between high sampling frequency data converters and high bit precision data converters. In addition in the mixed implementation case, since the delay spread asymptote is attained very quickly, the directive antennas can be used for reducing the delay spread of the channel (cf. Table VI) and thus profiting in a yielder manner of the high available operating frequency. The directive antennas allow to use very high operating frequencies, since the delay spread of the channel is reduced. Furthermore, the higher the working frequency is, the easier the directive antennas can be implemented. Thus there is an accenting phenomenon when the operating frequency increases.

As a result with the mixed techniques, thanks to higher achievable operating frequencies than in the mostly digital case, the channel capacity, in the mixed cases, can be higher than in the mostly digital case.

Table V exposes some state-of-art IR-UWB pulse generators. Table VI summarizes realistic IR-UWB channel delay spreads at 3-10 GHz and 60 GHz, for isotropic and directive antennas, extracted from the IEEE 802.15.4a and the 802.15.3c channel model. With these two kinds of information, we could determine some achievable data rates for mixed implementation transceivers. They are listed in table VII, binary and the M-ary modulations are also considered.

TABLE V

TABLE V
A SURVEY OF UWB PULSE GENERATOR CAPABILITIES.

| Year | Author | Technology | Pulse duration (ps) min | Pulse duration (ps) max | Ref |
|---|---|---|---|---|---|
| 2007 | Deparis et al. | pHEMT | 50 | 800 | [13] |
| 2007 | Badalawa et al. | CMOS 90 nm | 224 | - | [14] |
| 2006 | Kim et al. | CMOS | 380 | 4000 | [15] |
| 2006 | Bachelet et al. | CMOS 130 nm | 92 | - | [16] |

TABLE VI
RMS DELAY SPREAD FOR UWB CHANNEL AT 3-10 AND 60 GHZ IN FUNCTION OF ANTENNAS CONFIGURATIONS

| Residential LOS Channel | Half Power Beam Width Tx (°) | Half Power Beam Width Rx (°) | RMS Delay spread (ns) |
|---|---|---|---|
| UWB 3-10 GHz | 360 | 360 | 17 |
| UWB @ 60GHz | 360 | 360 | 7,718 |
| UWB @ 60GHz | 360 | 60 | 6,2 |
| UWB @ 60GHz | 360 | 15 | 3,455 |
| UWB @ 60GHz | 60 | 60 | 2,147 |
| UWB @ 60GHz | 60 | 15 | 0,948 |
| UWB @ 60GHz | 15 | 15 | 0,87 |

TABLE VII
SOME ACHIEVABLES VALUES OF DATA RATES, FOR IR-UWB MIXED IMPLEMENTATIONS. IMPACT OF THE CHANNEL DELAY SPREAD AND PULSE GENERATOR CAPABILITIES ARE EXPOSED

| | RMS Delay spread (ns) | Pulse Generator Ref. | Pulse Generator Bandwidth (GHz) | Channel capacity (Mbits/s) Binary Modulations | Channel capacity (Mbits/s) Ternary Modulations | Channel capacity (Mbits/s) M=4 Modulations |
|---|---|---|---|---|---|---|
| UWB | 17 | [15] | 2,63 | 57,54 | 115,07 | 172,61 |
| UWB | 17 | [14] | 4,46 | 58,06 | 116,12 | 174,17 |
| UWB | 17 | [16] | 10,87 | 58,51 | 117,01 | 175,52 |
| UWB 60 GHz | 7,718 | [16] | 10,87 | 128,04 | 256,08 | 384,12 |
| UWB 60 GHz | 6,2 | [16] | 10,87 | 158,93 | 317,86 | 476,80 |
| UWB 60 GHz | 3,455 | [16] | 10,87 | 281,93 | 563,86 | 845,79 |
| UWB 60 GHz | 2,147 | [16] | 10,87 | 446,63 | 893,26 | 1339,89 |
| UWB 60 GHz | 0,948 | [16] | 10,87 | 961,54 | 1923,08 | 2884,63 |
| UWB 60 GHz | 0,87 | [16] | 10,87 | 1039,51 | 2079,01 | 3118,52 |
| UWB 60 GHz | 0,87 | [13] | 20,00 | 1086,96 | 2173,91 | 3260,87 |

Table VII proves that the most important parameter is the delay spread of the channel, while the operating frequency is a second plan parameter.

## V. CONCLUSION

Starting from the Shannon channel capacity, we have exposed a new IR-UWB channel capacity based on the channel delay spread for the binary modulations over time variant multipath channels. This expression of the channel capacity is valid for a SNR greater or equal to 3dB. It's obtained under the no ISI assumption.

The mixed and mostly digital implementation impacts on the IR-UWB channel capacity are considered. Whatever the implementation, the channel delay spread is the main limitation. The channel capacity is bounded by a $1/d_{RMS}$ asymptote.

Concerning mostly digital radio, the sampling frequency of the data converters, as architecture key point, is used for evaluating the IR-UWB channel capacity. For high channel delay spreads the capacity is limited by the delay spread asymptote. I.e. a sampling frequency change doesn't impact significantly the achievable data rate. Whereas for low channel delay spread the sampling frequency impacts the capacity in a direct manner.

The same analysis concerning the delay spread asymptote and the importance of the operating frequency is done for the mixed implementations. However, in the mixed implementation case, the operating frequency values are severely larger than the state-of-the-art sampling frequencies of data converters. Due to this fact, the channel delay spread limitation is achieved more quickly than in mostly digital case. As a result in the mixed configurations, the channel capacity is almost totally dependent in channel delay spread. Increase the operating frequency is useless from a high data rate point of view. That's why we have exposed in this mixed case, the use of the M-ary modulations and the directive antennas for achieve higher capacity. The directive antennas reduce the channel delay spread.

At last, for a high data rate criteria comparison, mixed solution is more suited for two reasons. The M-ary modulations are not viable in mostly digital radio due to the ADC performances. The sampling frequency is drastically smaller than mixed operating frequency due to the $n_{sampling}$ factor (Shannon theorem).